 \documentclass[final,3p,times]{elsarticle}
\usepackage{graphicx}
\usepackage{amssymb}
\usepackage{subfig}
\usepackage{comment}
\usepackage{enumitem}
\usepackage{lineno}
\usepackage{hyperref}
\usepackage{xcolor}

\journal{Journal XYZ} %Name

\begin{document}

\begin{frontmatter}

% To Reopen or Not to Reopen Now?
% Feeling Like it is Time to Reopen Now?
%  Does USA Feel it is Time to Reopen Now?
\title{ \textit{Feeling Like It is Time to Reopen Now? \\ } COVID-19 New Normal Scenarios based on Reopening Sentiment Analytics}



\author[label1]{Jim Samuel}
\ead{jim@aiknowledgecenter.com}
\author[label2,label3]{Md. Mokhlesur Rahman}
\ead{mrahma12@uncc.edu}
\author[label1]{G. G. Md. Nawaz Ali\corref{cor1}}
\ead{ggmdnawazali@ucwv.edu}
\author[label4]{Yana
Samuel}
\ead{yana.samuel@gmail.com}
\author[label5]{Alexander Pelaez}
\ead{alexander.pelaez@hofstra.edu}
\address[label1]{University of Charleston, WV, USA}
\address[label2]{University of North Carolina at Charlotte, Charlotte, NC}
\address[label3]{Khulna University of Engineering \& Technology (KUET), Khulna, Bangladesh}
\address[label4]{Northeastern University, Boston, MA}
\address[label5]{Hofstra University, NY}
\cortext[cor1]{Corresponding author}

\begin{abstract}
The Coronavirus pandemic has created complex challenges and adverse circumstances. This research discovers public sentiment amidst problematic socioeconomic consequences of the lockdown, and explores ensuing four potential sentiment associated scenarios. The severity and brutality of COVID-19 have led to the development of extreme feelings, and emotional and mental healthcare challenges. This research identifies emotional consequences - the presence of extreme fear, confusion and volatile sentiments, mixed along with trust and anticipation. It is necessary to gauge dominant public sentiment trends for effective decisions and policies. This study analyzes public sentiment using Twitter Data, time-aligned to COVID-19, to identify dominant sentiment trends associated with the push to 'reopen' the economy. Present research uses textual analytics methodologies to analyze public sentiment support for two potential divergent scenarios - an early opening and a delayed opening, and consequences of each. Present research concludes on the basis of exploratory textual analytics and textual data visualization, that Tweets data from American Twitter users shows more trust sentiment support, than fear, for reopening the US economy. With additional validation, this could present a valuable time sensitive opportunity for state governments, the federal government, corporations and societal leaders to guide the nation into a successful new normal future. 
% In this paper, we are going to analyze the public sentiment about when the United States will go back to normal life after the COVID-19 pandemic. Over two months of lock down people get bored, jobless and frustrated. Now the peoples' ardent desire is when could we go back to our normal daily life.   
\end{abstract}


\end{frontmatter}

%\linenumbers
%\nolinenumbers
\section{Introduction}\label{Sect:Introduction}
\begin{center}
\textit{The anxiety, stress, financial strife, grief, and general uncertainty of this time will undoubtedly lead to behavioral health crises.} - Coe and Enomoto, McKinsey on the COVID-19 crisis \cite{Mckinsey_Returning}.
\end{center}

\begin{figure}[htbp]
    \centering
    \includegraphics[width=0.6\linewidth]{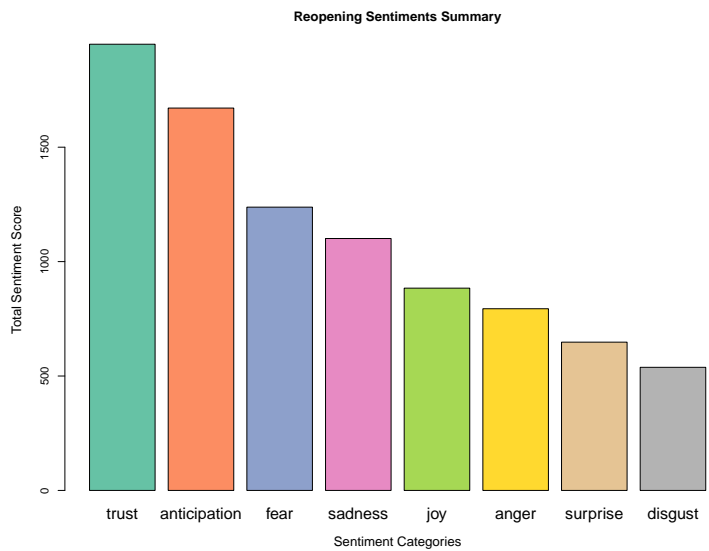}
    \caption{Reopening public sentiment graph.}
    \label{Fig:Reopneing_sentiment}
\end{figure}

COVID-19 has become a paradigm shifting phenomenon across domains and disciplines, affecting billions of people worldwide directly or indirectly. The current Coronavirus pandemic disaster has led to escalating emotional and mental health issues with significant consequences, and this presents a serious challenge to reopening and recovery initiatives \cite{goldmann2014mental}. In the United States (US) alone, COVID-19 has infected well over 1.25 million people and killed over 80,000, and continues to spread and claim more lives \cite{worldometers}. Moreover, as a result of the 'Lockdown' (present research uses Lockdown as the term representing the conditions resulting from state and federal government Novel Coronavirus response regulations and advisories restricting government, organizational, personal, travel and business functionalities), over 30 million people lost their jobs along with a multi-trillion dollar economic impact \cite{US_Labor}; furthermore these alarming numbers are growing everyday. There is tremendous dissatisfaction among common people due the continued physical, material and mental health challenges presented by the Lockdown, evidenced by the growing number of protests across the US \cite{Tension_over_restriction}.  There appears to be a significant sentiment, and a strong desire in people to go back to work, satisfy basic physical, mental and social needs, and eagerness to earn money as shown in exploratory public sentiment graph in Fig. \ref{Fig:Reopneing_sentiment}. However, there is also a significant sentiment to stay safe, and many prefer the stay-at-home Lockdown measures to ensure lower spread of the Coronavirus \cite{CDC_What_can_do}. While politicians may have vested interests, state governments and the federal government cannot ignore public sentiment. Therefore, to a consequential measure, the reopening of America, and its states and regions will be influenced by perceived public sentiment. 

Hence, it is important to address the question:  What is the dominant public sentiment in America concerning the reopening and in some form, going forward into a New Normal?. 'New normal' is a term being used to represent the potentially transformed socioeconomic systems \cite{Mckinsey_Beyond}, as result of COVID-19 issues, amidst the likelihood of multiple future waves of the Coronavirus pandemic. There is a fear that the current Coronavirus wave will see a spike in COVID-19 infections  with reopening and associated loosening of Lockdown restrictions. The present study analyzed public sentiment using Tweets from the first nine days of the month of May, 2020, with a keyword “reopen” from users with country denoted as USA, to gauge public sentiment along eight key dimensions of anger, anticipation, disgust, fear, joy, sadness, surprise and trust, as visualized in Fig. \ref{Fig:Reopneing_sentiment}. Twitter data and Tweets text corpus have been widely used in academic research and by practitioners in multiple disciplines, including education, healthcare, expert and intelligent systems, and information systems \cite{mohammadi2018academic, sinnenberg2017twitter, ghiassi2016targeted, lim2019mining, visvizi2019tweeting}. Sentiment analysis with Tweets presents a rich research opportunity, as hundreds of millions of Twitter users express their messages, ideas, opinions, feelings, understanding and beliefs through Twitter posts. Sentiment analysis has gained prominence in research with the development of advanced linguistic modeling frameworks, and can be performed using multiple well recognized methods, and tools such as R with readily available libraries and functions, and also through the use of custom textual analytics programming to identify dominant sentiment, behavior or characteristic trait \cite{saif2016contextual, samuel2014automating, pandey2017twitter}.  Tweets analytics have also been used to study, and provide valuable insights on a diverse range of topics from public sentiment, politics and pandemics to stock markets \cite{ansari2020analysis, samuel2020covid, kretinin2018going}. 
There are strong circumstantial motivations, along with urgent time-sensitivity, driving this research article. Firstly, in the US alone, we have seen over 89,000 deaths and over 1.4 million COVID-19 cases, at the point of writing this manuscript, and the numbers continue to increase. Secondly, there have been significant economic losses and mass psychological distress due to unprecedented job loss for tens of millions of people. These circumstantial motivations led us to our main research motivation: discovery of public sentiment towards reopening the US economy. The key question we seek to address is, what are the public sentiments and feelings about reopening the US economy?. Public sentiment trends insights would be very useful to gauge popular support, or the absence thereof, for any and all state level of federal reopening initiatives. 
\\

Scientists, researchers and physicians are suggesting that the reopening  process should be on a controlled, phased and watchful basis. The general recommendation is it should be done into three phases \cite{ReopenGuardian04_29}: \textit{Phase 1}: slow down the virus spread through complete Lockdown and very strict measures (for instance, mandatory stay-at-home orders, which have been active in many states). \textit{Phase 2}: reopen on a state-by-state, business-by-business, and block-by-block basis with caution. People would still be required to maintain social distancing, use PPE (personal protective equipment) and strict public hygiene \cite{Christopher2020Testing}. All of these would need to be implemented while protecting the most vulnerable sections of the population, using potential strategies such as the 'relax-at-home' or 'safe-at-home' mode. Moreover, the states would be expected to increase testing, tracing and tracking of COVID-19 cases, and ensure ample PPE supplies. \textit{Phase 3}: It would be possible for people to go back to near-normal pre-Coronavirus lifestyles, when proven vaccines and/or antibodies become commonly available, and mass-adoption sets in. Early research shows that 90\% of transmissions have taken place in closed environments, namely, home, workplace, restaurants, social gatherings and public transport, and such prevalence in critical social spaces contributed to increased fear sentiment  \cite{KnowRisk_Erin,Qian2020Indoor}. 
\\
\begin{figure}[htbp]
    \centering
    \includegraphics[width=0.55\linewidth]{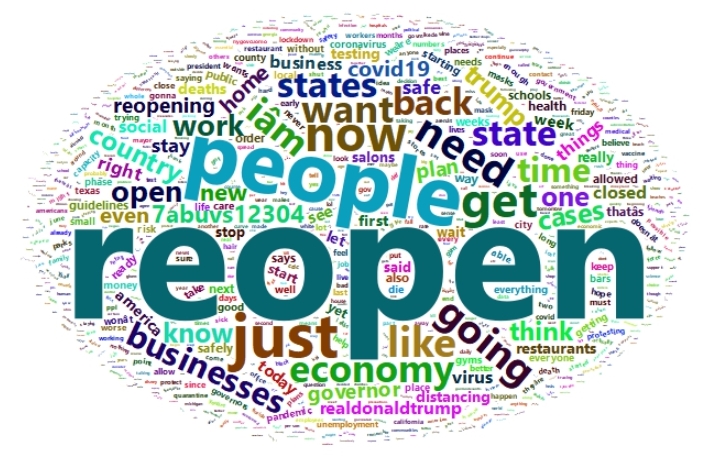}
    \caption{Word cloud about public sentiment.}
    \label{Fig:Reopening_wordcloud}
\end{figure}

From a current sentiments analysis perspective, since a large ratio of people need to start work, return to their businesses and jobs, and restart the economy, a quick reopening is perceived as being strongly desirable.  However, there is fear regarding a potential second outbreak of the COVID-19 pandemic, and this presents a cognitive dilemma with implications for mental health and emotional conditions. Information and information formats have an impact on human sentiment, behavior and performance, and it is possible to associate underlying feeling or belief to expressed performance and communication \cite{samuel2017information}. The present research addresses the COVID-19 public sentiment trend identification challenge by generating insights on popular sentiment about reopening the economy, using publicly available Tweets from users across the United States. These publicly downloadable Tweets were filtered with 'reopen' as the keyword, and insights were generated using textual data visualization, starting with the word cloud Figure \ref{Fig:Reopening_wordcloud} to gain a quick overview of the textual corpus, as such textual visualizations have the potential to provide both cross-sectional and narrative perspectives of data \cite{Conner2019picture}. Textual analytics were used to discover most frequently used words, phrases,  and prominent public sentiment categories. 

The present research is exploratory in nature and applied in its focus. Given the urgency of the COVID-19 situation in the United States and worldwide, this study has reluctantly forfeited the intellectual luxury of time consuming adoption and integration of theoretical frameworks, which would satisfy academic rigor but would forego contributing practically to critically time-sensitive and desperately needed COVID-19 recovery and reopening solutions. Consequentially, this research is part of our COVID-19 solutions research stream, which has enthusiastically embraced the generation of insights using rapid exploratory analytics and logical induction, with timely and significantly practical implications for policy makers, local, state and federal governments, organizations and leaders to understand and prepare for sentiment-sensitive scenarios. This research has thus aimed to discover the most critical insights, optimized to time constraints, through sentiment-sensitive reopening scenarios analysis. The rest of the paper is organized as follows. Section \ref{Sect2:Scenario_analysis} highlights the scenario analysis of past and current events and public sentiment. Section \ref{Sect3:Methods} demonstrates the adopted method for analyzing public sentiment. Section \ref{Sect4:Discussion} is dedicated for in-depth discussion, pointing out limitations and opportunities of this research. Finally, Section \ref{Sect5:Conclusion} concludes this paper. 

% Roopa and Asha predict diabetes disease with 82\%  accuracy with a simple linear model \cite{Roopa2009Linear}. They use PCA (Principle Component Analysis) for extracting features of diabetes and then apply the linear regression method for forming the prediction model. 
%  We should not have references with xy author said so & so - we must have our own arguments and then cite supporting references. 
%

\section{Scenario Analysis: COVID-19 Sentiment Fallout}\label{Sect2:Scenario_analysis}

Extant research has illustrated the dramatic growth in public fear sentiment using textual analytics for identifying dominant sentiments in Coronavirus Tweets \cite{samuel2020covid}. Public fear sentiment was driven by a number of alarming facts as described in the subsections below. To start with, we list a number of initiatives aimed at understanding, explaining and predicting various aspects of the Coronavirus phenomena. A number of works made early prediction of the spread of COVID-19 \cite{Zhong2020Early,Aboelkassem2020Pandemic}. Zhong et al. \cite{Zhong2020Early} made the early prediction of the spread of the coronavirus using simple epidemic model called SIR (Susceptible-Infected-Removed) dynamic model. The model works with two key parameters; the infection rate and the removal rate. The infection rate is the number of infections by one infective by unit time and the removal rate is the ratio of the removed number to the number of infectives, where the removed number includes recovered and dead infectives. Y. Aboelkassem \cite{Aboelkassem2020Pandemic} used a Hill-type mathematical model to predict the number of infections and deaths and the possible re-opening dates for a number of countries. The model works based on the three main estimated parameters; the steady state number (the saturation point), the number of days when the cases attain the half of the projected maximum, and Hill-exponent value. %These parameters are estimated using  some optimization techniques and already available data for a country. 
Li et al. \cite{Li2020Characterizing} categorized COVID-19 related social media post into seven situational information categories. They also identify some key features in predicting the reposted amount of each categorical information. The predicting accuracy of different supervised learning methods are different: the accuracy of SVM (Support Vector Machine), Naive Bayes, and Random Forest are 54\%, 45\% and  65\%, respectively. 

\subsection{COVID-19 people sentiment-impact: Fear}

%COVID-19 illness and hospitalization 
\begin{figure}[htbp]
    \centering
    \subfloat[COVID-19 Emergency room visits.]{\includegraphics[width=0.5\linewidth]{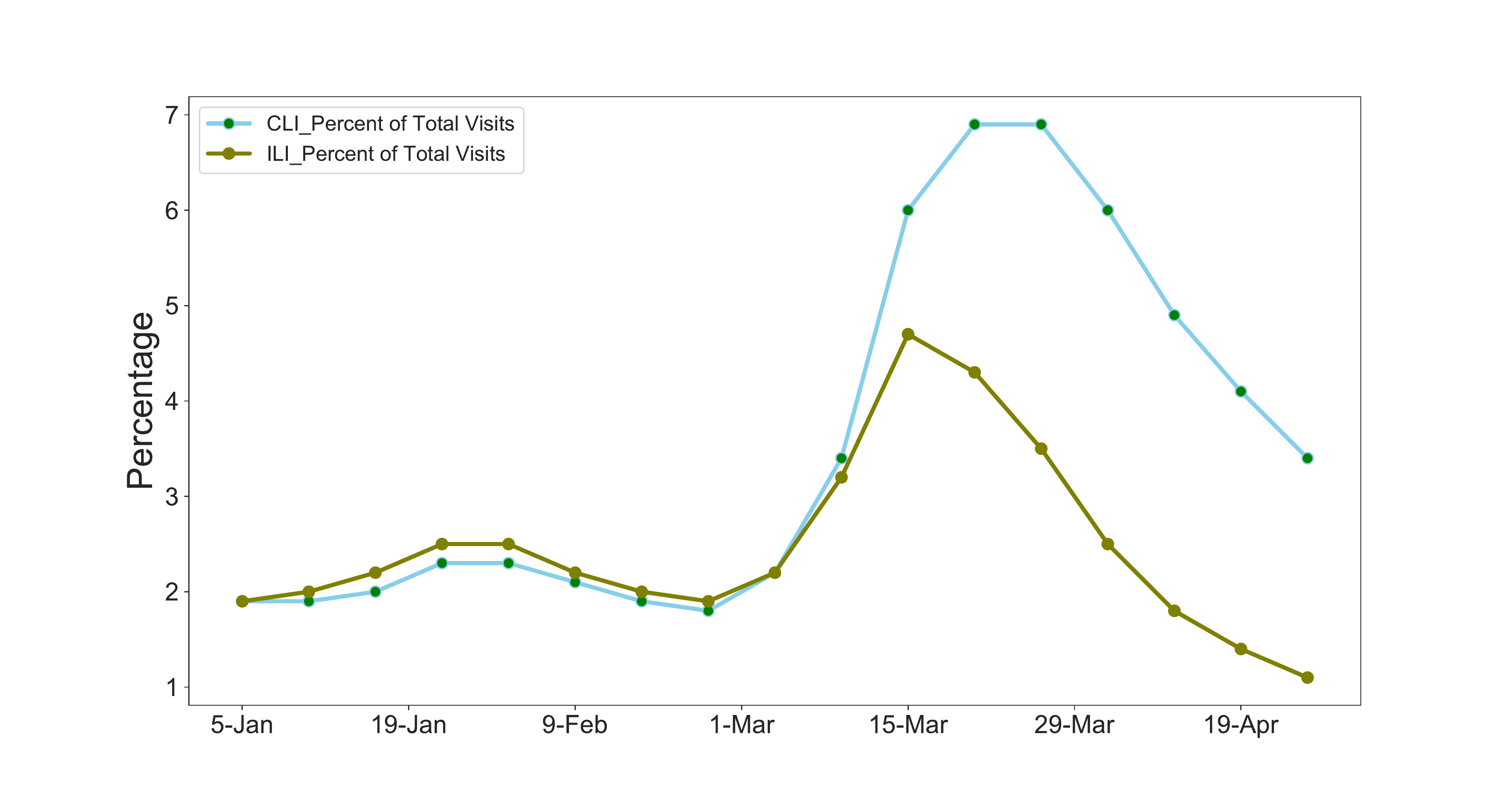}\label{Fig:COVID_illness}}
    %\par
    \subfloat[Hospitalizations per 100,000 of population.]{\includegraphics[width=0.5\linewidth]{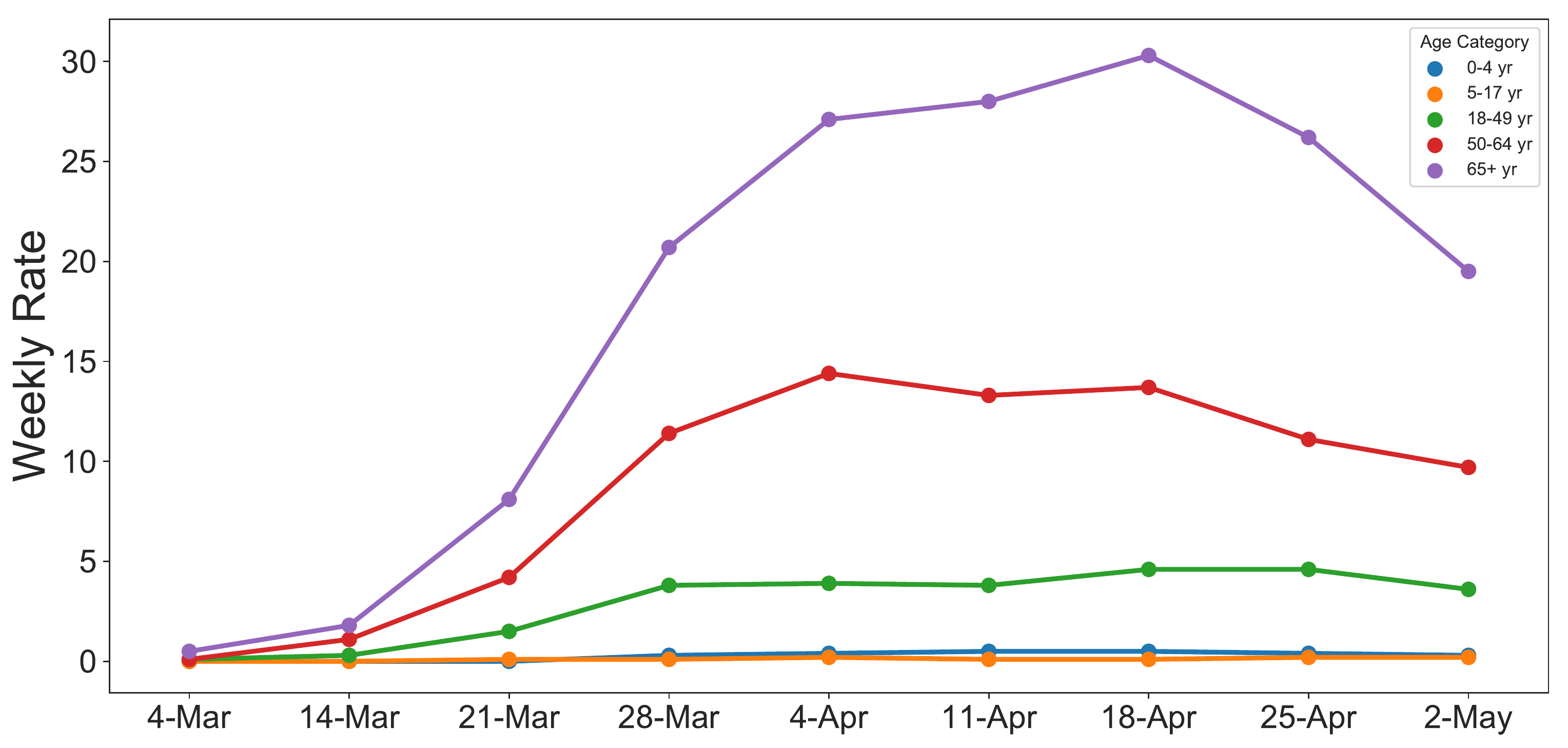}\label{Fig:Hospitalization}}
    \caption{Impact of COVID-19 outbreak on medical resources in US (source, CDC \cite{CDC_CLI_ILI}).}
    \label{Fig:COVID_illness_hospitalization}
\end{figure}

In the US, concurrent to the physical healthcare problem, extant research observes the mass fear sentiment about Cornonavirus and COVID-19 phenomena to be on a significant growth curve from the time the study started tracking the sentiment, in the February  and climbing steeply towards March, 2020 \cite{samuel2020covid}. According to Center for Disease Control and Prevention (CDC), this was also around the time that a massive number of people started seeking medical attention for COVID-19 conditions \cite{CDC_CLI_ILI}. Fig. \ref{Fig:COVID_illness} shows the percentage of patients visits for COVID-19-like illnesses (CLI) and Influenza-like illnesses (ILI), compared to the total number of emergency department visits from December 1, 2019 to April 26, 2020 in the Unites States (source, CDC \cite{CDC_CLI_ILI}). It clearly indicates that a significant number of people began using emergency facilities for CLI and ILI from around March of 2020, and this corresponds with the growth fear and panic sentiments associated with COVID-19. However, by late April, we see the emergency visits curve relaxing along with a decline in the number of new infections in many states in the US. Fig. \ref{Fig:Hospitalization}  exhibits COVID-19-associated weekly hospitalizations per 100,000 of the US population among different age categories from March 04 to May 02, 2020 in the US. It shows that from the second week of March, a significant number of people aged over 65 years needed to be hospitalized; and the hospitalizations count peaks by the middle of April. It clearly shows that this age group is amongst the most vulnerable to the COVID-19 outbreak. The age group of 50-64 years follows as the second most impacted, and the age group of 18-49 years follows as being the third most impacted. However, COVID-19 had a very limited impact on the 0-17 years age group. Healthcare experts and researchers continue to monitor, develop and apply solutions that are helping physical recovery. From a current mental healthcare and sentiment tracking perspective, we did not find any reports highlighting changes to early stage COVID-19 fear and panic sentiment, nor clear updates on sentiment trends from extant literature. Hence, having identified an important gap, this present research identifies changes in public sentiment by collecting and analyzing Twitter data for the first part of May 2020, as described in the methods section. 

%Cases_infected_despair maps 
\begin{figure}[htbp]
    \centering
    %\subfloat[Infected cases.]{\includegraphics[width=0.5\linewidth]{Figures/Cases_map.pdf}\label{Fig:Cases_map}}
    \subfloat[Death cases.]{\includegraphics[width=0.5\linewidth]{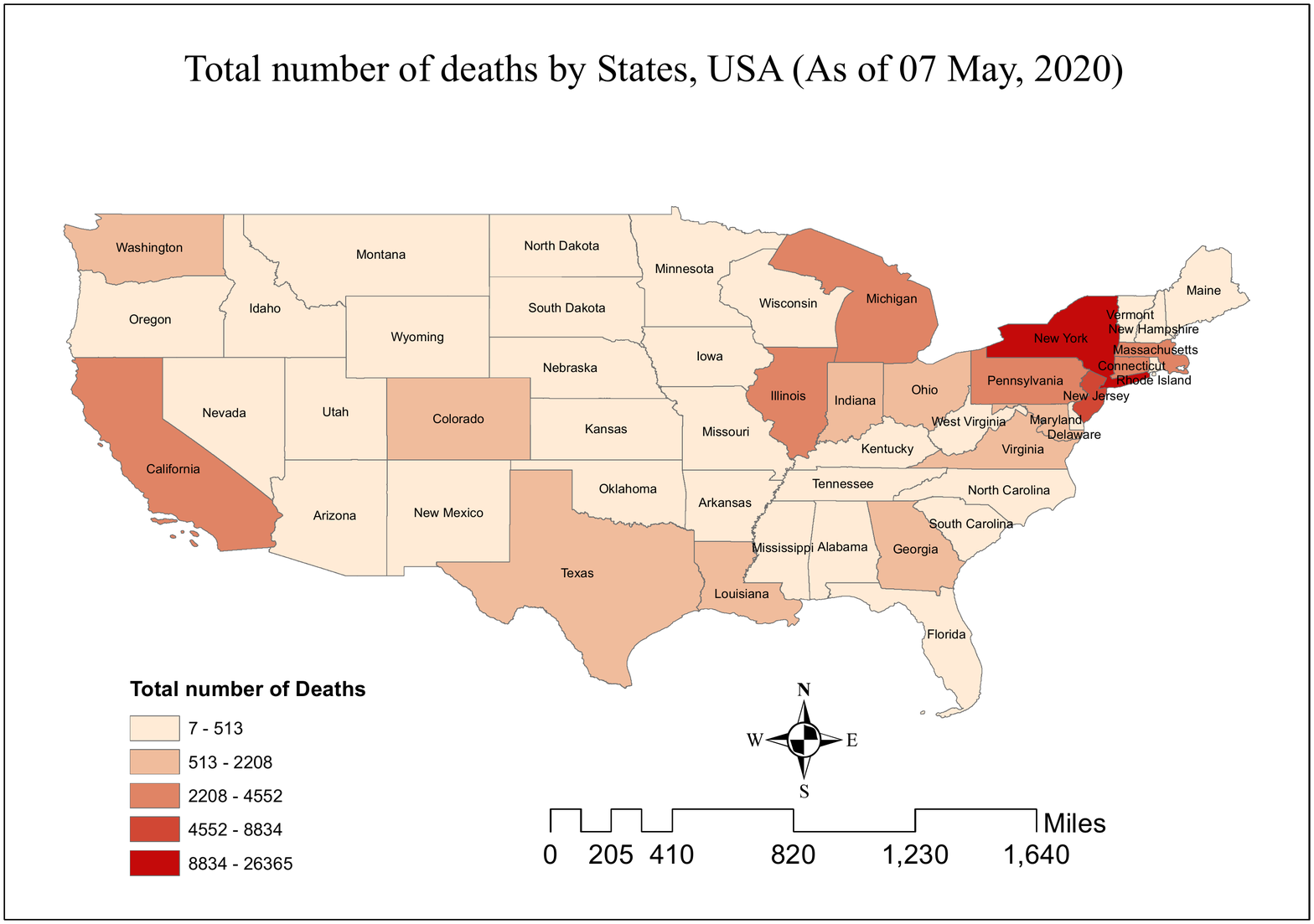}\label{Fig:Death_map}}
    %\par
    \subfloat[Deaths of despair.]{\includegraphics[width=0.5\linewidth]{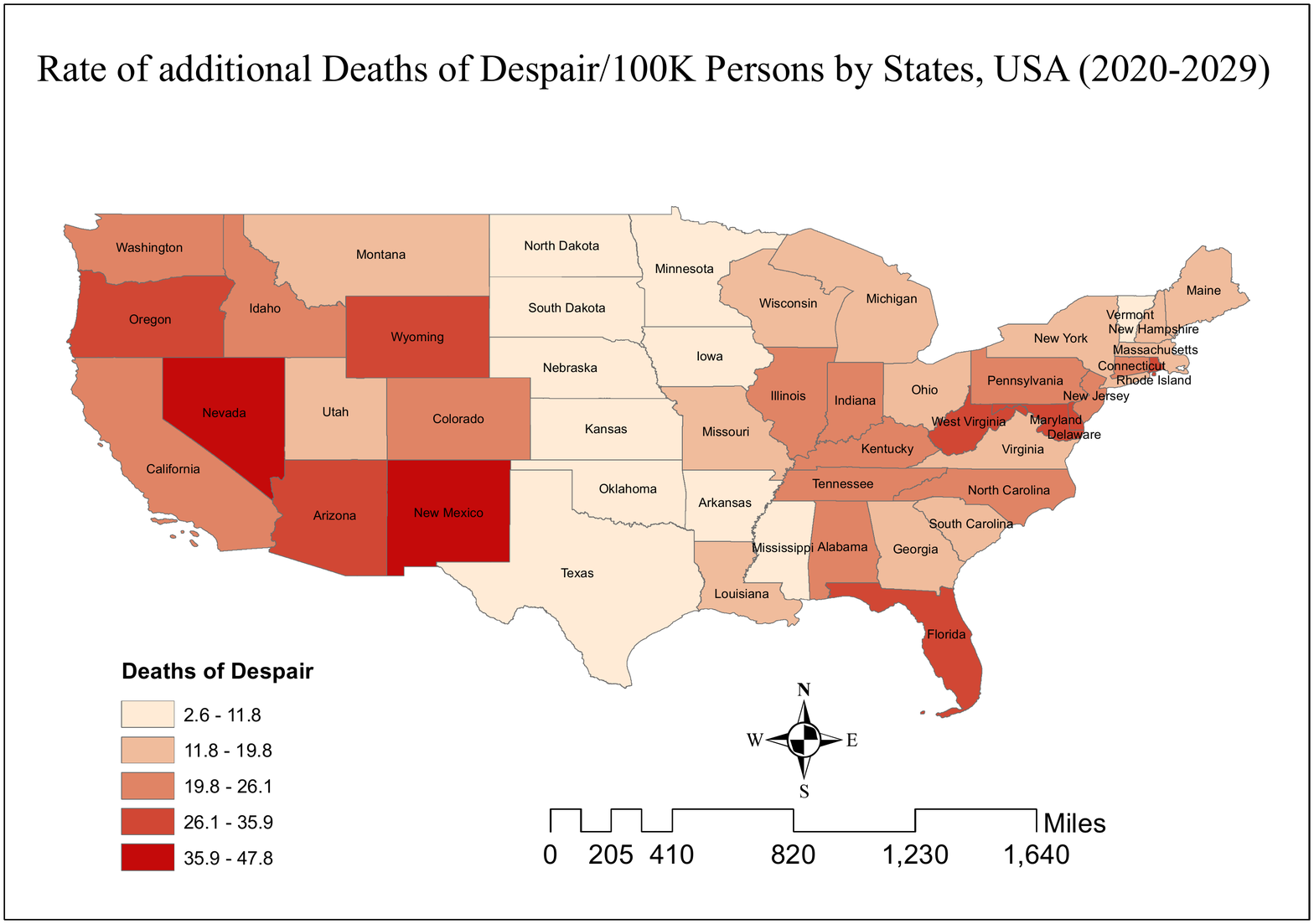}\label{Fig:Death_of_despair_map}}
    \caption{COVID-19 outbreak by states as of May 7, 2019 (source, worldometers \cite{worldometers}).}
    \label{Fig:cases_death_despair_map}
\end{figure}

\subsection{Societal sentiment-impact: Despair}  

\textit{The whole is often greater than the sum of its parts} - this adage holds true regarding the adverse collective psychological, sentimental and mental healthcare impact of the Coronavirus and COVID-19 phenomena on human society and societal structures. While there was a significant growth in fear and anxiety on the individual level, collectively as a society, these took the shape of panic and despair, as evidenced in panic-buying driven shortages of items in super markets for which there were no supply chain disruptions, indicating that these shortages were driven by adverse public sentiment. Certain American states and regions were more drastically impacted than the others. COVID-19 had a severe impact on New York and New Jersey, followed by Massachusetts, Illinois, Michigan and California as shown in Fig. \ref{Fig:cases_death_despair_map} (source, worldometers \cite{worldometers}). So far New York and New Jersey have seen the  maximum number of deaths (\ref{Fig:Death_map}). Interestingly, if we notice the death of despair prediction for 2020-2029 as shown in Fig. \ref{Fig:Death_of_despair_map}, New York and New Jersey are not in the most affected list, instead some states, which perhaps have less job opportunity and higher older population, will suffer the most (source, Well Being Trust \& The Robert Graham Center \cite{Death_of_despair}). The prediction shows that the most likely states to bear a negative impact in the long run will be New Mexico, Nevada, Wyoming, Oregon, Florida and West Virginia, due to the looming socioeconomic fallout of COVID-19. This implies a longer term and more complex mental healthcare challenge, as states begin their journey to recovery - it will be vital for governments and relevant organizations to track, understand and be sensitive to shifts in public sentiment at the local and national levels. 

%Unemployment 
\begin{figure}
    \centering
    \includegraphics[width=0.7\linewidth]{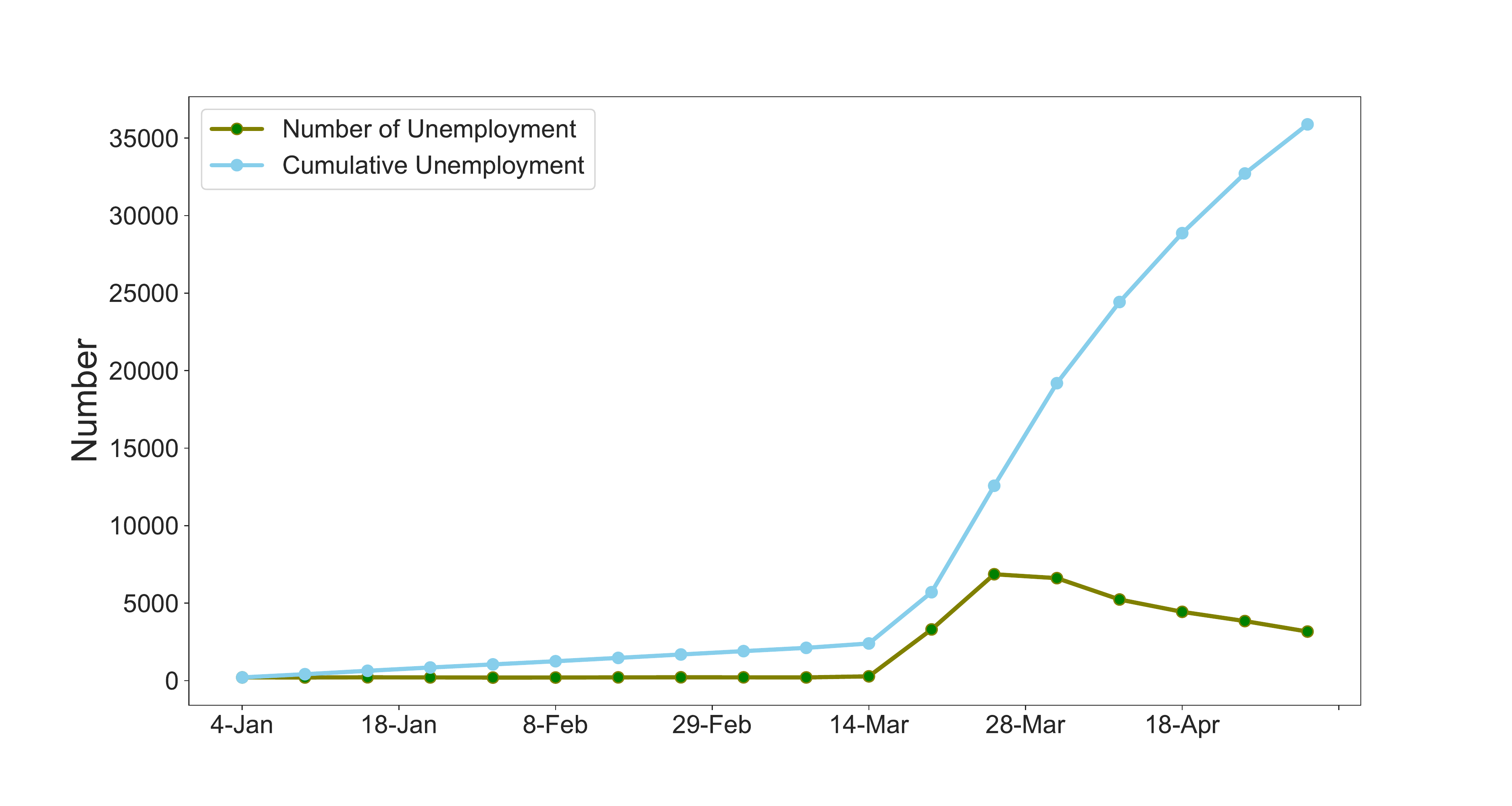}
    \caption{Unemployment situation in US due to COVID-19 (source, US Labor \cite{US_Labor}).}
    \label{Fig:Unemployment}
\end{figure}

\begin{comment}
\begin{figure}[htbp]
\centering
\subfloat[Unemployment rate.]
{\includegraphics[width=0.5\linewidth]{Figures/Unemployment_Rate} 
 \label{Fig:Unemployment_rate}}
%\par
\subfloat[Unemployment count.]
{\includegraphics[width=0.5\linewidth]{Figures/Unemployment_number} 
 \label{Fig:Unemployment_count}}
 \par
\subfloat[Unemployment by states.]
{\includegraphics[width=0.55\linewidth]{Figures/Unemployment_map.pdf} 
 \label{Fig:Unemployment_map}}
\caption{Unemployment situation in US due to COVID-19 (source, US Labor \cite{US_Labor}).}
\label{Fig:Unemployment}
\end{figure}
\end{comment}

\subsection{Economic-downturn sentiment-impact: Confusion}

% labor supply which results in the 
The Coronavirus pandemic has caused significant global disruptions, leading to continuing losses valued at trillions of US dollars due to the closure of many businesses and entire industries, such as hospitality and air travel. However, confusion is rampant due to counter efforts by governments, nationalized monetary policy interventions, and large scale financial assistance to individual citizens, organizations and businesses. For example, the US government has successfully provided multi-trillion dollar stimulus checks, small business assistance and a wide range of concessions and equity market support mechanisms. Such initiatives make it difficult for investors, and individuals to gauge fundamental value of many assets as government policy intervention can support systemic market shifts, and disrupt forecasts. Risks in the global financial market have elevated substantially due to investors' anxiety in this pandemic crisis and assets price variability \cite {coronavirusworld, papadamou2020direct, zhang2020financial}. Data analysis on daily COVID-19 cases and stock market returns for 64 countries demonstrated a strong negative return with increasing confirmed cases \cite{ashrafstock}. The stock market reacted strongly in the early days of confirmed cases compared to a certain period after the initial confirmed cases. In contrast, the reaction of individual stock markets is closely related to the severity of the local outbreaks, which leads to economic losses, high volatility and unpredictable financial markets, and Tweets textual analytics have been used to gauge associated sentiment \cite {zhang2020financial, kretinin2018going}. All of these COVID-19 effects lead to a delicate market equilibrium, immersed in a fair degree of confusion, positively supported by expectation of government intervention, and limited by the negative consequences of COVID-19.

Also, COVID-19 has caused a severe interruption in the functioning of businesses and institutions, leading to loss of employment, diminished income opportunities and disruption of labor markets, leading to significant increase in distress \cite {adams2020inequality}. A prolonged pandemic may lead to mass unemployment and irreversible business failures \cite {zhang2020financial}. According to the International Labor Organization (ILO), the global unemployment rate will increase by 3\%$\sim$13\%, subsequently it will increase the underemployment and reduce the economic activities such as global tourism \cite{figari2020welfare}. Adams-Prassl et al., \cite {adams2020inequality} studied the comparative job losses and found that about 18\% and 15\% of people lost their jobs in early April in the US and UK, respectively, whereas only 5\% people lost their job in Germany. They also predicted that the probability of losing the job of an individual is about 37\% in the US and 32\% in the UK and 25\% in Germany in May. Evidently, people are worried and anxious about their livelihood and future income, driving a sense of urgency for reopening the economy. Fig. \ref{Fig:Unemployment} demonstrates how the  unemployment situation in the US is worsened by the COVID-19 pandemic (source, US Department of Labor \cite{US_Labor}). By the mid of March, 2020, many people started losing their jobs and by late April around 16\% of the population became unemployed. %(\ref{Fig:Unemployment_rate}). 
As of now over 30 million people have lost their jobs, and this number continues to grow as the Lockdown continues. %(\ref{Fig:Unemployment_count}). %If we see state-wise, the most people lost their jobs in those states where coronavirus hit the most, for instance, job wise most sufferer states are New York, California, Texas, Michigan, Pennsylvania, Georgia and Florida (\ref{Fig:Unemployment_map}).  
The global economic downturn caused by COVID-19 is estimated to be worst since the great recession in 2008 \cite {figari2020welfare}. COVID-19 related shutdown is expected to affect 20\%-25\% output decline in advanced economies (e.g., US, Canada, Germany, Italy, French) with a 33\% drop in consumer expenditure, and around 2\% decline in annual GDP growth rate \cite {oecd2020initialimpact}. It is projected that annual global GDP growth will be reduced to 2.4\% in 2020 with a negative growth rate in the first quarter of 2020, from a weak rate of 2.9\% in 2019 \cite {coronavirusworld}. Compared to the pre-COVID-19 period, it is estimated that about 22\% of the US economy would be threatened, 24\% of jobs would be in danger and total wage income could be reduced by around 17\% \cite {del2020supply}.        
%Consequently, it is deadly and adversely affecting people, economy, industrial production, supply chain, financial markets, and travel in different parts of the world \cite {yu2020modelling, zhang2020financial, coronavirusworld}. 

\section{Method and Current Sentiment Analytics}\label{Sect3:Methods}
Thus far, this study has used secondary data and extant research to motivate, inform and direct the research focus towards current sentiment analytics on the subject of reopening the US economy. Previous sections have summarized key aspects of the extensive socioeconomic damage caused by the Coronavirus pandemic,  and highlighted associated psychological and sentiment behavior challenges. For the purposes of the main data analysis, this research uses a unique Twitter dataset of public data specifically collected for this study using a custom date range and filtered to be most relevant to the reopening discussion. While public sentiment has changed from apathy, disregard and humor in the earliest stages of the Coronavirus pandemic, to fear in February and March of 2020, and despair in March and April of 2020, yet there is a lack of clarity on the nature of public sentiment surrounding reopening of the economy. The analysis of Twitter data uses textual analytics methods that includes discovery of high frequency key words, phrases and word sequences that reflect public thinking on the topic of reopening. These publicly posted key words, phrases and word sequences also allow us to peek into the direction of evolving public sentiment. Anecdotal Tweets provide insights into special cases and they provide a peek into influential Tweets and logical inflection points. In the final parts of the data analysis, the study provides insights into dominant current sentiment with sentiment analysis using the R programming language, associated sentiment analysis libraries (R packages) and lexicons.   
%Table 
\begin{table}[htbp]
\centering
\caption{Twitter data features: Mentions \& Hashtags.}\label{Table:endogenous_features}
%Mention counts
\subfloat[Mention count.]{
\begin{tabular}{ll}
\hline
\multicolumn{1}{l}{\textbf{Tagged}} & \multicolumn{1}{l}{\textbf{Rank}} \\ \hline
realDonaldTrump   & 1        \\
GovMikeDeWine     & 2        \\
NYGovCuomo        & 3        \\
GavinNewsom       & 4        \\
GovLarryHogan     & 5         \\
\hline             
\end{tabular}}
\hspace{1.5cm}
%Hashtag count
\subfloat[Hashtag count.]{
\begin{tabular}{ll}
\hline
\textbf{Hashtag}                        & \textbf{Rank} \\
\hline
COVID19                                & 1                  \\
BREAKING                               & 2                  \\
Covid\_19                              & 3                  \\
coronavirus                            & 4                  \\
Texas                                  & 5                  \\
\hline
%BREAKING abc13   hounews               & 2                 
\end{tabular}}
\end{table}
\subsection{Method} \label{Method}
The present study used Twitter data from May 2020 to gauge sentiment associated with reopening. 293,597 Tweets, with 90 variables, were downloaded with a date range from 04/30/2020 to 05/08/2020 using the rTweet package in R and associated Twitter API, using the keyword “reopen”. This follows standard process for topic specific data acquisition and the dataset was saved in .rds and .csv formats for exploratory data analysis \cite{pepin2017visual, samuel2020beyond}. The complete data acquisition and analysis were performed using the R programming language and relevant packages. The dataset was cleaned, filtered of potential bot activity and subset by country to ensure a final subset of clean Tweets with country tagged as US. It is possible that the process omitted other Tweets from the US which were not tagged by country as belonging to the US. Furthermore, for ethical and moral purposes, we used custom code to replace all identifiable abusive words with a unique word text "abuvs” appended with numbers to ensure a distinct sequence of characters. While we wanted to avoid the display of abusive words in an academic manuscript, we still believed it to be useful to retain the implication that abusive words were used, for further analysis. After the filtering, cleaning and stopwords processes, a final dataset consisting of 2507 Tweets and twenty nine variables was used for all the Twitter data analysis, textual analytics and textual data visualization in this paper. A visual summary of the key words in the textual corpus of the filtered Tweets is provided in the word cloud visualization in Fig. \ref{Fig:Reopening_wordcloud}.   

\subsubsection{N-Grams Word Associations}\label{Subsect:Word_Association}
\begin{figure}[htbp]
    \centering
    \includegraphics[width=0.85\linewidth]{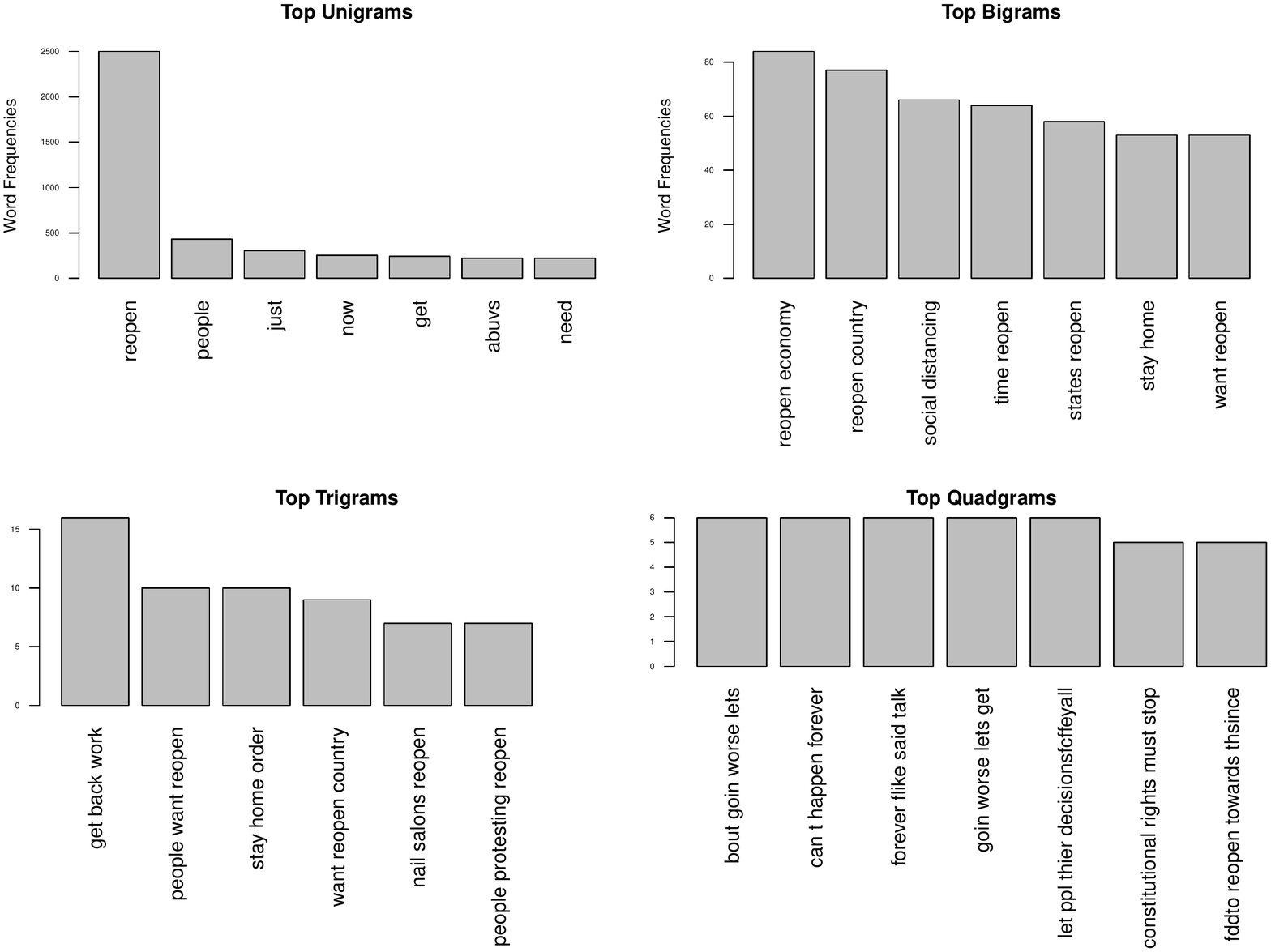}
    \caption{N-Grams.}
    \label{Fig:N_grams}
\end{figure}

The text component of the dataset, consisting of filtered Tweets only was used to create a text corpus for analysis and data visualization. Word frequency and N-grams analysis were used to study the text corpus of all the tweets, to discover dominant patterns, and are summarized in Fig. \ref{Fig:N_grams}. Word frequency analysis revealed an anticipated array of high frequency words including economy, states, businesses, COVID, open, back, work, country, reopening, plan and governor. N-grams, which focus on the identification of frequently used word pairs and word sequences revealed interesting patterns. The most frequent Bigrams (two word sequences) included: open economy, reopen country, social distancing, time reopen, states reopen and want reopen. These largely indicate positive sentiment towards reopening. The most frequent Trigrams (three word sequences) included: get back work, people want reopen, stay home order and want reopen country. These trigrams also indicate medium to strong support for the reopening. The most frequent "Quadgrams" (four word sequences) included: can’t happen forever, goin worse lets get, and constitutional rights must stop. Quadgrams reveal more complex sentiment, with positive but weak support for reopening. For example, ‘can’t happen forever’  most likely implies that the Lockdown cannot last indefinitely, and ‘constitutional rights must stop’ most likely implying that intrusive Lockdown measures are not appreciated.  

%ngrams 

\subsubsection{Descriptive Analysis of Tweets}\label{Subsect: Describe}
\begin{figure}[htbp]
    \centering
    \includegraphics[width=0.75\linewidth]{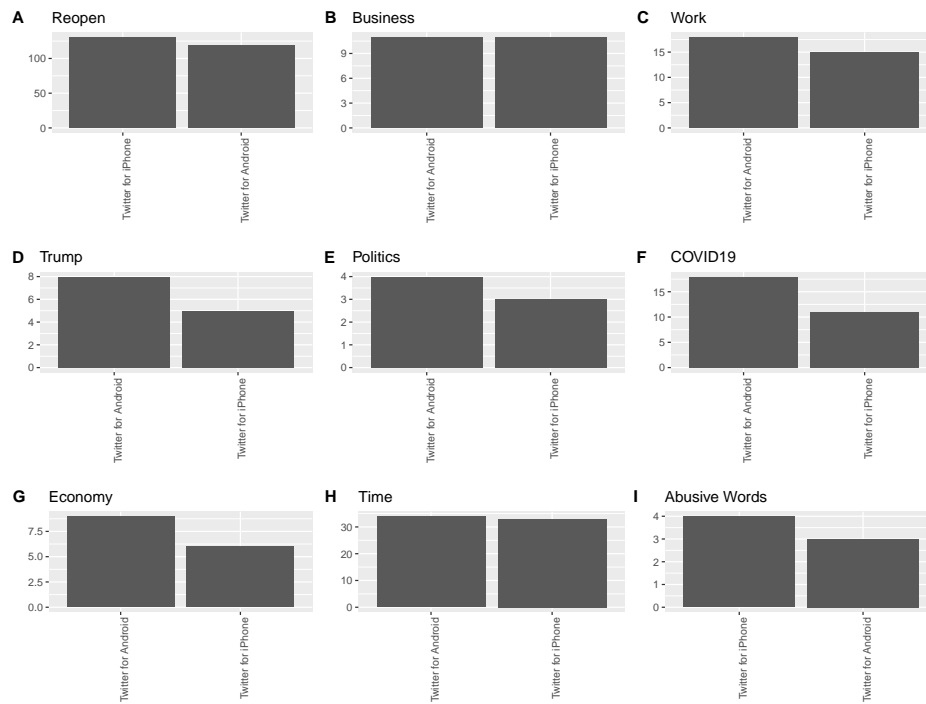}
    \caption{Tweets grouped by device type}
    \label{Fig:AndroidvsIphone}
\end{figure}

Descriptive analysis was used to explore the data and Tables \ref{Table:endogenous_features} and \ref{Tab:Locations} summarize the reopening Twitter data features associated with sentiment analysis. Table \ref{Table:endogenous_features}(a) ranks the dominant Twitter user names mentioned in the reopening Tweets data, and Table \ref{Table:endogenous_features}(b) ranks the leading hash tags used in the data. The text of the Tweets was also analyzed in conjunction with other variables in the dataset to gain insights into behavioral patterns, which included grouping by technology (device used to post Tweet) and analysis of key words usage by such technological classification. We grouped Tweets into two technology-user classes: iPhone users and Android users, to explore potential behavioral aspects. Extant research support such a classification of technology users for analyzing sentiment, psychological factors, individual differences and technology switching \cite{miller2012smartphone, samuel2020covid, shaw2016individual, lin2014understanding}. We identified 1794 Twitter for iPhone users, and 621 Twitter for Android users, in our dataset and ignored smaller classes, such as users of web client technologies. For the purposes of relative analysis, we normalized the technological groupings so as to ensure comparison of ratio intrinsic to each group, and to avoid the distortion caused by unequal number of users in the two technology-user classes. Our analysis and grouped data visualizations revealed some interesting patterns as summarized in Fig. \ref{Fig:AndroidvsIphone}. Twitter for iPhone users had a marginally higher ratio of ‘reopen’ mentions, while they were at par with Twitter for Android users in their references to ‘business’ and ‘time’ urgency words. Twitter for iPhone users tended to use more abusive words, while Twitter for Android users tended to post more Tweets referencing ‘work’, ‘Trump’, ‘Politics’, ‘COVID-19’ and ‘economy’. Negative sentiment is usually tagged to the use of abusive words, positive sentiment is currently associated with reopening words and either positive or negative sentiment could be associated with political, work, business and economy words, subject to context and timing. This reveals that, technology user groups may differ in their sentiment towards reopening and that by itself could merit additional research focus. 

%Android_vs_Iphone

\subsubsection{Illustrative Tweets and Sentiment}\label{Subsect: Anecdote}

\begin{center}
\textit{"Reopen everything now!!!!"}  5/3/2020, 21:46 Hrs, Twitter for iPhone\\
\textit{"NO state is ready to reopen"} 5/2/2020, 23:55 Hrs, Twitter for Android\\
\end{center}
It is important to connect textual analytics discoveries, such as word frequencies and N-grams sequences, to context and demonstrate potential ways in which the popular words and word sequences are being used. To that purpose, this section presents and discusses a few notable Tweets associated with the N-grams and descriptive textual analytics. Name mentions, abusive words and hashtags in the Tweets displayed in this paper have been deleted or replaced to maintain confidentiality, research ethics and relevance, and spellings, though incorrect have been retained as posted originally. We observed Tweets with high positive sentiment and emotional appeal such as “What a beautiful morning to reopen the economy!” and “Ready to use common sense and reopen our country”. Some of the Tweets focused on humor: “More importantly when do the bars reopen”, “Day 50: I find it amusing that skinheads are the ones calling to reopen hair salons”, “The first weekend the bars reopen is going to kill more people than the coronavirus” and “First karaoke song when the bars reopen will be a whole new world”. A large number of Tweets referenced jobs, work and businesses, such as: “Then tell the states to reopen. That is the only way to create jobs”, “Then tell the states to reopen. That is the only way to create jobs”, “How about you just reopen the state?", “WE ARE BEGGING YOU TO OPEN UP THE ECONOMY BUT YOU DONT CARE! Our jobs won’t be there if you keep this going!” and “I don't want to get it but we must all reopen and get back to work”. 
%4 types sentiment
\begin{figure}[htbp]
    \centering
    \includegraphics[width=0.75\linewidth]{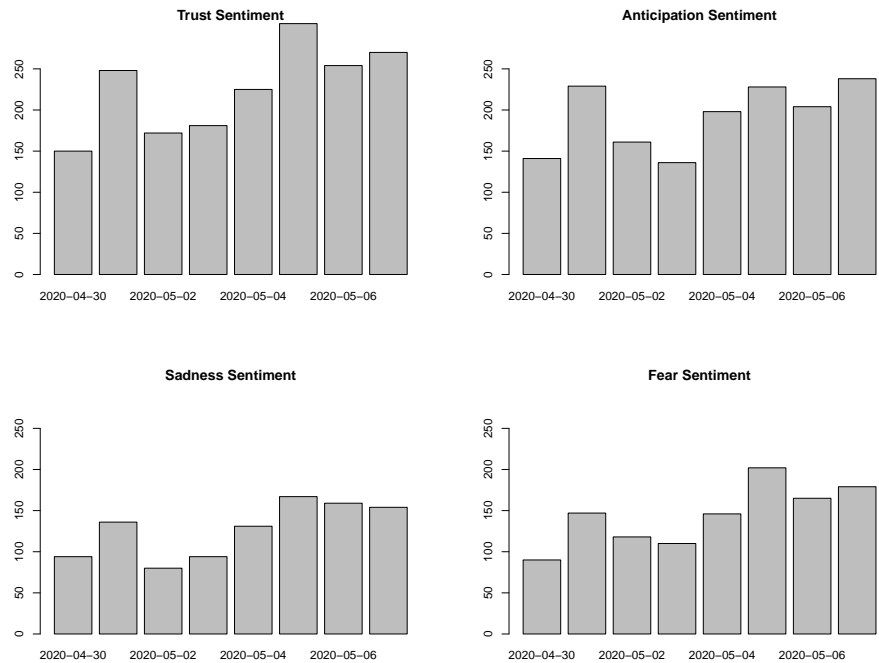}
    \caption{Different type sentiments as time progresses.}
    \label{Fig:4_sentiments}
\end{figure}

Many Tweets were political: “History will also show that during the pandemic the Democrats did nothing to reopen the country or economy but continued to collect a paycheck while 30\% was unemployed”, “Trump thinks he needs the states to reopen for his reelection” and celebrities were not spared of the negative sentiment: “Melinda Gates, multi-billionaire hanging out comfortably at home, insists America not reopen the economy until at least January and until America implements paid family medical leave. \#outOfTouch” and “Bill Gates laughs when he hears about the economy falling because he wants you to die”. Some Tweets expressed skepticism: “This is not hard. The economy won't get better even if you open up EVERYTHING because consumer consumption is based on CONFIDENCE. Economics 101 y'all. America ain't ready to reopen”, “The need to reopen the economy is definitely evidence that capitalism will kill us all” and "What happens when the 20\% win and stores reopen and the other 80\% still refuse to show up?”. Frustration was evident in some Tweets “i never want to hear the words ‘reopen’ and ‘economy’ ever again”, and some Tweets appealed to logic: “The cure can’t be worse than the virus. It’s time to reopen America”, “If they haven't been preparing by now, it's their problem. Many others have spent their time getting ready to Reopen”, “But we don’t have proof that will happen. When is a good time to reopen? It will always be risky” and “More will be devistated if we don't reopen. Follow the protocol set out and get us back to work”. Also, some quoted caution: “I do believe there will be serious soul searching in a few weeks, as states reopen and coronavirus case numbers explode”, while other Tweets emphasized individual rights: “It is really past time to reopen our country and to allow US citizens our constitutional rights” and “Reopen. Let owner make a living. No one is being forced To go there. Bring back choice”. As evident, many strong, complex and diverse emotions are expressed in these Tweets examples, and it is nearly impossible to manually estimate cumulative sentiment classes or scores for large Tweets datasets. However, with the development of standardized sentiment scoring and classification technologies, it has become efficient to perform sentiment analysis on large and unstructured data, and current research leverages R tools to perform sentiment analysis on reopening data. 

\subsection{Sentiment analysis}\label{Subsect: sentir}
\begin{figure}[htbp]
\centering
\subfloat[Sentiment Score.]
{\includegraphics[width=0.5\linewidth]{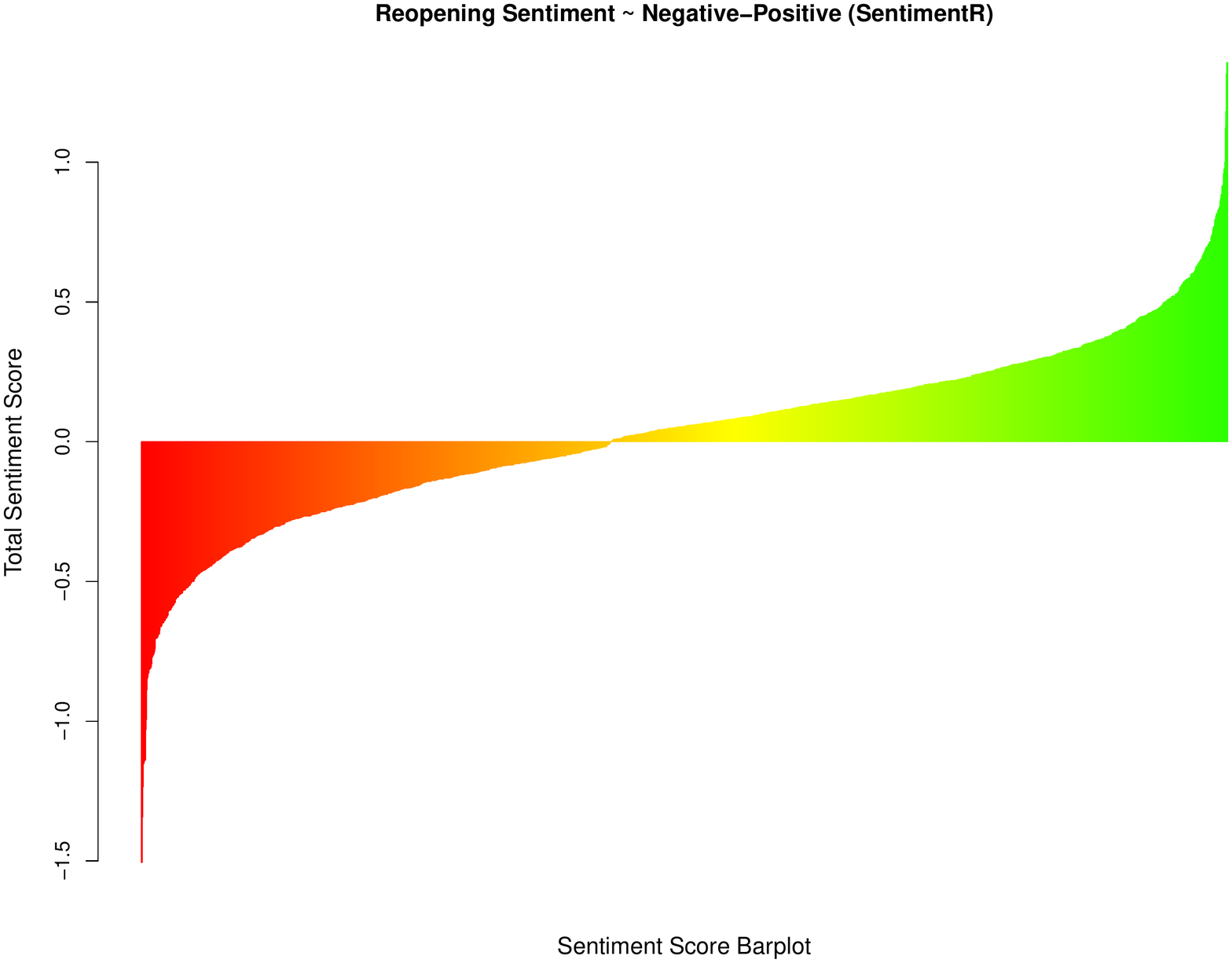} 
 \label{Fig:PosNegBarplot}}
 %\par
\subfloat[Sentiment Class.]
{\includegraphics[width=0.5\linewidth]{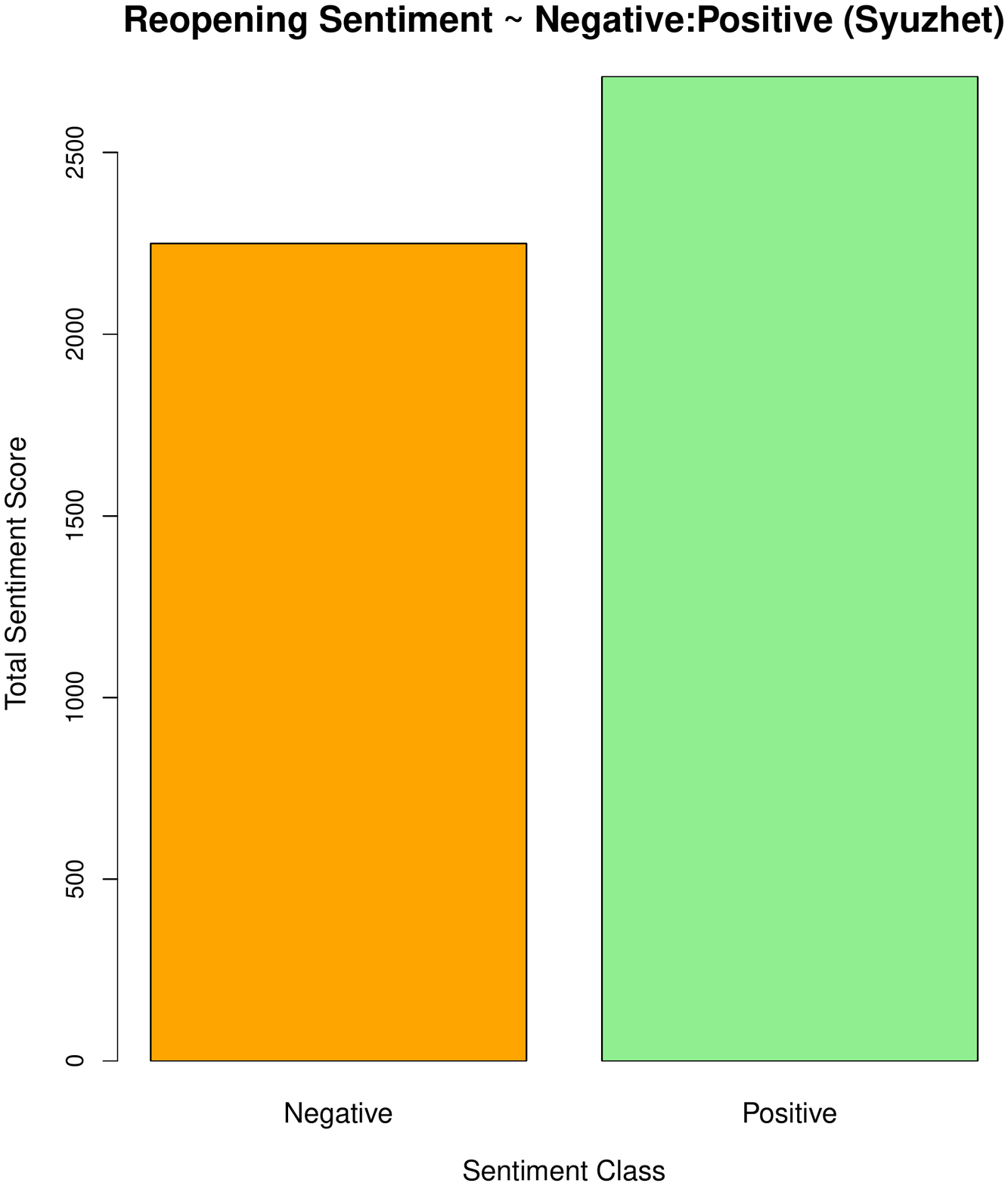} 
 \label{Fig:BarpPN}}
\caption{Positive and Negative sentiment about reopening.}
\label{Fig:pos_neg}
\end{figure}

The scaling up of computing technologies over the past decade has made it possible for vast quantities of unstructured data to be analyzed for patterns, including the identification of human sentiment expressed in textual data. Sentiment analysis is one of the main research insights benefits from textual analytics as it extracts potentially intended sentiment meaning from the text being analyzed. Early stage past research used custom methods and researcher defined protocols to identify both sentiment and personality traits such as dominance, through the analysis of electronic chat data, and standardized methods to assign positive and negative sentiment scores \cite{ samuel2014automating,he2015novel,ravi2015survey,samuel2018going}. Sentiment analysis assigns sentiment scores and classes, by matching keywords and word sequences in the text being analyzed, with prewritten lexicons and corresponding scores or classes. For this research, we used R and well known R packages Syuzhet and sentimentr to classify and score the reopening Tweets dataset \cite{R_Syuz, R_senti}. The R package Syuzhet was used to classify the Tweets into eight sentiment classes as shown in Fig. \ref{Fig:Reopneing_sentiment}. Syuzhet also measures positive and negative sentiment by a simple sum of unit positive and negative values assigned to the text, and therefore a single complex enough Tweet may simultaneously be scored as having a positive score of 2 and a negative score of 1. Sentir measures words and associated word sequence nuances, providing a score ranging from around -1 to 1, and the values could be less than -1 or greater than 1. The final sentiment score from sentiment package is a summation of sentiment values assigned to parts of the sentence (or textual field) and can be less than -1 or more than 1, as shown in Fig. \ref{Fig:pos_neg}. An analysis of the reopening Tweets displayed 48.27\% of positive sentiment, 36.82\% of negative sentiment and 14.92\% of neutral sentiment, with sentiment. The sentiment analysis, combing scores from sentiment and the classification provided by Syuzhet highlighting trust and anticipation reflected a largely positive sentiment towards reopening the economy. 

%pos vs neg sentiment
\begin{comment}
\begin{figure}[htbp]
    \centering
    \includegraphics[width=0.49\linewidth]{Figures/PosNeg.pdf}
    \caption{Postive and Negative sentiment about reopening.}
    \label{Fig:pos_neg}
\end{figure}
\end{comment}

\subsubsection{ Sentiment Scenario Analysis}  
There is a high level of uncertainty regarding future events, and questions surrounding the effectiveness of available healthcare facilities in protecting the populace against a potential second wave of Coronavirus remain. There is a serious concern that an unfettered and undisciplined reopening of the US economy may lead to rapid spreading of the Coronavirus and a corresponding steep increase in COVID-19 cases and deaths. However, it is also clear that states cannot keep businesses and services closed indefinitely. This situation needs to be addressed from multiple perspectives, and there are numerous efforts underway to develop solutions for phased openings and preparation for the New Normal. The present study seeks to leverage the insights discovered through timely sentiment analytics of the reopening Tweets data, and apply the findings to the "Open Now", including planned and phased openings, versus "Open (indefinitely) later", including advocacy of maintaining complete shutdown, by analyzing four potential New Normal scenarios:

\begin{enumerate}[label=\alph*)]
\item \textbf{Scenario 1:} Positive public sentiment trend and reopen now
\item \textbf{Scenario 2:} Positive public sentiment trend and reopen later
\item \textbf{Scenario 3:} Negative public sentiment trend and reopen now
\item \textbf{Scenario 4:} Negative public sentiment trend and reopen later 
\end{enumerate}    

These New Normal scenarios are highlighted and discussed on a \textit{ceteris-paribus} basis, that is we only consider sentiment variance and reopening timing, holding all else equal, such as the progression of COVID-19, healthcare and social distancing protocols, and all other necessary precautions, preparations and “ongoing intensive deep sanitization cycles, physical-distance protocols and associated personal and community paraphernalia" \cite{Samuel8Principles}.

\subsubsection{Positive sentiment scenarios a. \& b.}	
Extant research has demonstrated the validity of using Tweets sentiment analysis to understand human behavior \cite{ibrahim2019decoding}. Sentiment analysis of the reopening Tweets data has indicated dominant sentiment trends for trust and anticipation, and a larger proportion of positive sentiment. Scenario \textbf{a} is a valuable New Normal setting from a leadership and policy perspective, as it provides the enthusiasm and public support that is required for the massive efforts that will be needed to wind the economy back up into action. A positive and supportive public attitude will also help the government, and public and private organizations in implementing health protocols more effectively than if positive sentiment were missing. Scenario \textbf{b} in contrast will be a missed New Normal opportunity, where in spite of positive public sentiment trends, there is failure to reopen in a timely manner. There are risks associated with any decision or strategy that could be applied, however, the risk of losing the support of positive public sentiment is too significant to be ignored. There is a likelihood, that the positive sentiment and forward looking energy to reopen, restart businesses, rejoin work and push ahead, may be dominated by fear, panic and despair once again due to prolonged socioeconomic stress and financial loss. It will remain important for societal leaders and responsible agencies to seize positive sentiment trends indicating support for reopening, and make the most of it, especially in the absence of deterministic healthcare and socioeconomic solutions.

\subsubsection{Negative sentiment scenarios c. \& d.}
If negative public sentiment were to dominate, then it would have the potential to hinder both a quick reopening, and any effective reopening at any later stage. Many Tweets have expressed extreme negative sentiment, such as ``[Abusive word] people reopen the whole freaking country and let everybody die ..." and ``we are all going to die". Interestingly sentiment analysis showed that, though the number of positive sentiment Tweets were higher, the negativity of the negative sentiment score Tweets were more extreme than the positiveness of the positive sentiment score Tweets. The extreme negative Tweet had a sentiment score of $\sim$1.51, and the extreme positive sentiment score was $\sim$1.36, implying that though fewer Tweets were negative, the intensity of their linguistic expression was higher. Scenario \textbf{c} would lead to a volatile start to the New Normal scenario, as it would lead to a reopening without adequate public sentiment support. Reopening now with the absence of a dominant positive public sentiment trend could generate a number of adverse effects, including failure of businesses that attempt to reopen. It is still possible that some quick reopening associated successes and positive information flows, reopening now under scenario \textbf{c} may still lead to a limited measure of New Normal success as the economy restarts and businesses and individuals are empowered to create economic value. However, scenario \textbf{d}, \textit{ceteris-paribus}, has the potential to lead to the most adverse New Normal scenarios. Scenario \textbf{d} is a combination of dominant negative public sentiment trends, where sentiment categories such as fear and sadness dominate, and negative sentiment scores peak along with a delayed opening without time constraints. Indefinite Lockdown has never been an option, and the strategy has been to flatten the curve through Lockdown and distancing measures, so as to create a buffer time zone for healthcare facilities to prepare and cater to the infected, and for governments to make plans and preparations for optimal post-reopening New Normal scenarios. Scenario \textbf{d} has the potential to lead to growth in negative public sentiment trends, creating long term mental healthcare challenges, coupled with rapidly deteriorating economic fundamentals. 

In summary, current reopening data analytics indicate a positive public sentiment trend, which allows for the the more favorable outcomes \textbf{a} and \textbf{b}, which can lead to optimal New Normal scenarios which maximize the benefits of reopening, while limiting related risks. 

%Key word frequency 
% \begin{figure}[htbp]
%     \centering
%     \includegraphics[width=0.75\linewidth]{Figures/KeyWordFreq.pdf}
%     \caption{Key word frequency.}
%     \label{Fig:Key_Word_Freq}
% \end{figure}

\section{Discussion: Limitations, risks and opportunities.} \label{Sect4:Discussion}
The present research has a few limitations, which must be kept in mind for any application of the public sentiment trends insights and New Normal scenarios presented in this research. These limitations also present opportunities for future studies and further extension of this research. Post-COVID-19 reopening and recovery are complex challenges with significant uncertainties and unknowns - this is a new systemic crisis for which the world has no established solutions or proven models to depend on. Needless to say, any reopening endeavor or the absence thereof, any action or inaction, are all fraught with significant risks. We identify and discuss some of these risks from a sentiment association perspective. The subsections below provide elaborations on the limitations of this research, risks and opportunities for future work.  

%Tagged Location table 
\begin{table}[htbp]
\centering
\caption{Twitter data features: Locations.}\label{Tab:Locations}
\subfloat[Tagged.]{\begin{tabular}{ll}
\hline
\multicolumn{1}{l}{\textbf{Location}} & \multicolumn{1}{l}{\textbf{Rank}} \\ \hline
Los Angeles, CA     & 1     \\
Brooklyn, NY        & 2     \\
Chicago, IL         & 3     \\
Florida, USA        & 4     \\
Pennsylvania, USA   & 5     \\
Manhattan, NY       & 6     \\
North Carolina, USA  & 7    \\
Houston, TX         & 8     \\
San Francisco, CA   & 9     \\
\hline
\end{tabular}}
\hspace{1.5cm}
\subfloat[Stated.]{\begin{tabular}{ll}
\hline
\multicolumn{1}{l}{\textbf{Location}} & \multicolumn{1}{l}{\textbf{Frequency}} \\ \hline
Los Angeles, CA     & 1        \\
United States       & 2        \\
Brooklyn, NY        & 3        \\
New York, NY        & 4        \\
Las Vegas, NV       & 5        \\
Orlando, FL         & 6        \\
Chicago, IL         & 7        \\
Dallas, TX          & 8        \\
North Carolina, USA & 9       \\
\hline
\end{tabular}}

\end{table}

\subsection{Limitations}
There are two areas of limitations of this research: quality of data and variance in sentiment analysis tools. Twitter data has been used extensively in research and in practice. However, the data is susceptible to bot activity and also to data errors caused voluntarily or inadvertently by users who provide inaccurate data. For example, Table \ref{Tab:Locations} (a) \& (b) are both location variables, and apart from the fact that they are poorly tagged, they do not provide a reliable interpretation of actual location of the user. Extensive cleaning and data preparation are necessary, which is often time and resource consuming, especially with textual data. Furthermore, though large volumes of public Twitter data can be acquired with reasonable effort, the quality of data can be affected by bot activity, repeat posts and spam posts. Though the reopening Tweets data used for the present research was cleaned and well prepared prior to analysis, following standard processes, yet the likelihood that the algorithmic processes did not successfully address all issues remains. Secondly, the tools available for sentiment analysis are subject to a measure of error, and are subject to the scope of the underlying lexicons. That is also the reason why using multiple lexicons could lead to somewhat different results depending on the context and the complexity of the underlying textual corpus \cite{S2019Viral}. This limitation in sentiment analysis tools can usually be mitigated by analyzing a larger number of Tweets. This research was intended to be exploratory and directional in nature and therefore, multiple data sources were not used. Ideally, public sentiment must be gauged through multiple listening mechanisms, using data from multiple social media platforms and public communications, to provide a better representation of the population for sentiment analysis.

\subsection{Reopening Risks}
Fear became a prominent public sentiment as awareness of the seriousness and devastating effects of the Coronavirus pandemic increased \cite{samuel2020covid}. This sentiment was not unjustified, due to diverse risks associated with the pandemic. COVID-19 has a higher transmissibility with a reproduction number (R0) of 2.0 - 6.47 (average R0 is 3.58) which indicates that the disease can be transmitted to 2.0 - 6.47 people from an average infected person \cite {liu2020reproductive,yu2020modelling}. This transmissibility rate is higher than recent infection diseases such as SARS (Severe Acute Respiratory Syndrome) and Ebola which have a reproduction number of 2 – 5 \cite {liu2020reproductive}. Hence, COVID-19 is highly contagious, especially within enclosed spaces such as trains, buses, restaurants, crowded factory floors, indoor markets and similar spaces. However, it is also well known that COVID-19 mainly impacts only the vulnerable parts of the population (commonly known as those with preexisting conditions and the elderly with weak immune systems), and this awareness has led to growing concerns and protests for reopening businesses and workplaces around the world. It is a public health issue to assess the safety of workplaces, and estimate their likelihood to transmit contagious diseases rapidly through a variety of activities (e.g., customer and patient dealings, close contact interaction with colleagues) \cite {edwards2016influenza, webster2019systematic, baker2020burden}. For example, healthcare workers (90\% are exposed more than once a month and 75\% are exposed more than once a week) bear a higher risk of getting infected, and thus may constitute a sub-segment for sentiment analysis, which can aid mental health evaluation. Besides, some other occupations (e.g., police, firefighters, couriers, social workers, daycare teachers, and construction workers) have a higher number of exposed workers in the United States. Self-isolation can significantly reduce ICU beds requirements, and flatten the disease outbreak curve. With a R0 of 2.5 and without self-isolation, 3.8 times the number of ICU beds would be required in the US to treat critically affected people \cite {moghadas2020projecting}. In contrast, about 20\% of self-isolation by infected persons, could reduce ICU beds by 48.4\%. With a RO of 2, self-isolation can reduces ICU bed requirements by 73.5\%. Knowledge of infectious disease transmission in workplaces, social distancing and stay at home practices are critical safeguards from rapid spread of infections \cite {edwards2016influenza}. Thus, for reopening workplaces and sustaining the economy, it is crucial to adopt appropriate protective measures (i.e., PPE, mandatory influenza symptoms sick leave) besides adequate workplace settings (i.e., emergency preparedness, risk mitigation plans, personal hygiene) to reduce risk and spreading of COVID-19 \cite {dryhurst2020risk, moghadas2020projecting}. Sentiment analysis can help track and manage public sentiment, as well as local or groups sentiments, subject to availability of suitable and timely data, and thus contribute to risk mitigation. 

% Boston et al., (2020) \cite {bostan2020assessments} observed that 31.7\% of the healthcare workers had contact with COVID-19 cases in Turkey. About 27.3\% of the healthcare workers who provided services patients are diagnosed with COVID-19. 

To prevent rapid transmission of COVID-19, most of the affected countries around the world implemented varying forms of Lockdown policies (e.g., quarantine, travel ban, social distancing), with significant economic consequences \cite {moser2020pandemic, zhang2020financial, yu2020modelling}. During this Lockdown period, people were forced to stay at home, and many lost their jobs, leading to significant emotional upheavals. Two recent surveys demonstrated that numerous small businesses have shut down due to COVID-19 \cite {bartik2020small, metlife2020small}. Collecting data from 5,800 small businesses, Bartik et. al. \cite {bartik2020small} found that about 43\% of them are temporarily closed, and on average they reduced their employee counts by 40\% as compared to the beginning of the year. According to the US Chamber of Commerce, about 24\% of all small businesses are already closed, and another 40\% of those who have not closed are  quite likely to close by June of 2020. Despite these alarming numbers, about 46\% of those surveyed believe that within a period of six to twelve months, the US will return to a state of normalcy \cite {metlife2020small}. About 25\% of all workers in the United States who are employed in information technology, administration, financial, and engineering sectors can perform work from home \cite {baker2020cannot}. In contrast, about 75\% of workers employed in healthcare, industrial manufacturing, retail and food services, etc. are unable to perform work from home due to the nature of thier work which mandates physical presence. Consequently, a large portion of the workers (108.4 M) are at risk of adverse health outcomes. Reopening the economy will exacerbate the situation with a higher potential for COVID-19 transmission. Not reopening the economy has the potential to create irreversible socioeconomic damage which can subsequently lead to greater loss of lives and more pain, along with diminished long term capabilities to fight the Coronavirus pandemic, and other crises, should they persist of arise. Sentiment analysis, and public sentiment trends based scenario analysis can inform and enlighten decision makers to make better decisions and thus help mitigate risks in crisis, disaster and pandemic circumstances.

\subsection{Opportunities}
Technology supported sentiment analysis presents numerous opportunities for value creation through serving as a support mechanism for high quality decision making, and for increasing awareness leading to risks mitigation. For researchers and academicians, this study presents a new sub-stream of research which provides a way to use sentiment analysis in crisis management scenario analysis. Extant research has demonstrated that information facets and categories can influence human thinking and performance, and hence can impact feelings. This presents an opportunity for future research to perform sentiment analysis with multiple information facets at varying levels to examine potential behavioral variations \cite{samuel2017effects}. This study also provides pandemic specific public sentiment insights to researchers specifically interested in pursuing the Coronavirus pandemic and COVID-19 studies. Academicians can expand the current study using additional sentiment analysis tools and customized crisis relevant sentiment analysis lexicons. There is also an easy opportunity for researchers to gain rapid insights by repeating this study for future time periods and thus help develop a sequence of reopening and recovery relevant public sentiment analysis studies. Such an approach would also be very useful for practitioners who have a strong need for recognizing market sentiment for a wide range of business decisions. 

Practitioners can make use of this study in at least two ways: the first is a direct application of the positive public sentiment exploratory findings of this study, after additional validation and due diligence, to inform and improve organizational decisions; and the second is to utilize the logic of methods presented in this research for situation specific and localized sentiment trends based scenario analysis related to organizational decisions. For example, the four New Normal scenarios schema can be customized and adapted to business situations, where modified sentiment estimation methods can be used to gauge specific consumer segment sentiment and used to inform products (or services) design, manufacturing (or process) and marketing decisions. The [positive sentiment : negative sentiment] X [action 1 (now) : action 2 (later)] matrix, from the `Sentiment Scenario Analysis' section provides a conceptual scenario analysis framework which has high potential for scaling and adaptation.

\section{Conclusions and Future Work} \label{Sect5:Conclusion}
\begin{center}
\textit{``We shall draw from the heart of suffering itself\\
the means of inspiration and survival."} – Churchill \cite{Samuel8Principles} 
\end{center}

Sentiment analysis is a valuable research mechanism to support insights formation for the reopening and recovery process. Sentiment analysis tools can be effectively used to gauge public sentiment trends from social media data, especially given the high levels of trust in social media posts among many networks \cite{shareef2020group}. This study discovered positive public sentiment trends for the early part of May, 2020. Specifically, the study discovered high levels of trust sentiment and anticipation sentiment, mixed with relatively lower levels of fear and sadness. Positive sentiment count dominated negative sentiment count, though negative Tweets used more extreme language. While sentiment analysis provides insights into public feelings, it is only one part of a complex set of contributions which will be required to effectively move into the New Normal. Sentiment analysis based on the Tweets data used in this study, appears to indicate public sentiment support for reopening. The message that ``it is time to reopen now" was prominent in the reopening Tweets data. Furthermore, we believe that this research contributes to the sentiment component of the recent call for cyberpsychology research, and well developed sentiment analysis tools and models can help explain and classify human behavior under pandemic and similar crisis conditions \cite{guitton2013developing, guitton2020cyberpsychology}. Positive public support sentiment will remain critical for successful reopening and recovery, and therefore additional initiatives tracking sentiment and evaluating public feelings effectively will be extremely useful. Sentiments can be volatile and localized, and hence technological and information systems initiatives with real-time sentiment tracking and population segmentation mechanisms will provide the most valuable public sentiment trends insights. With additional research and validation\footnote{This is a pre-print version of an under review Journal article and some portions may be reduced /absent, including the statistical /quantitative analysis in Section \#3}, the research strategy utilized in this paper can be easily replicated with necessary variations, and applied on future data from multiple social media sources, to generate critical and timely insights that can help form an optimal New Normal.



%% References



%% New version of the num-names style
%\bibliographystyle{elsarticle-num-names}
%\bibliographystyle{elsarticle-num}
\bibliographystyle{elsarticle-num}
\bibliography{Bibliography.bib,COVID_Analytics_MDPI.bib}

\begin{comment}
\appendix {Note:} 
\footnote{This is a pre-print version of an under review Journal article and some portions may be reduced /absent, including the statistical /quantitative analysis in Section \#3}
\textit{\textbf{[This is a pre-print version of an under review Journal article and some portions may be reduced /absent, including the statistical /quantitative analysis in Section \#3 ]}}
\end{comment}





% \end{thebibliography}

\end{document}